\documentclass{aip-cp}

\usepackage[numbers]{natbib}
\usepackage{rotating}
\usepackage{graphicx}
\usepackage{url}

\begin{document}

\title{The Extended Jet In AP Librae As The Source Of The VHE $\gamma$-ray Emission}

\author[aff1,aff2]{Michael Zacharias\corref{cor1}}
\author[aff1]{Stefan J. Wagner}

\affil[aff1]{Landessternwarte, Universit\"at Heidelberg, K\"onigstuhl, D-69117 Heidelberg, Germany}
\affil[aff2]{Now at: Centre for Space Research, North-West University, Potchefstroom 2520, South Africa}
\corresp[cor1]{E-mail: m.zacharias@lsw.uni-heidelberg.de}

\maketitle

\begin{abstract}
Most modeling attempts of blazars use a small emission zone located close to the central black hole in order to explain the broad-band spectral energy distribution. Here we present a case where additionally to the small region a $>$kpc-scale jet is required to successfully reproduce the spectrum and especially the TeV emission, namely the low-frequeny peaked BL Lac object AP Librae detected in the TeV domain by the H.E.S.S. experiment. Given that other parts of the spectral energy distribution follow the characteristics implied by the source classification, the inverse Compton component spans 10 orders of magnitude, which cannot be reproduced by the one-zone model. Additionally, observational constraints in both the synchrotron and inverse Compton compoenent strongly constrain the parameters of a self-consistent model ruling out the possibility of TeV photon production in the vicinity of the galactic center. We discuss the possibility that the TeV radiation is emitted by highly energetic particles in the extended, arcsec-scale jet, which has been detected at radio and X-ray energies. The slope of the jet X-ray spectrum indicates an inverse Compton origin, and an extrapolation to higher energies coincides with a break feature in the $\gamma$-ray band. Modeling the jet emission with inverse Compton scattering of the cosmic microwave background results in an excellent fit of the radio, X-ray and TeV emission. Implications will be discussed, such as properties of the jet, acceleration scenarios, and observations to test the model. If confirmed, large scale jets are able to efficiently accelerate particles and to keep relativistic speeds up to distances of several 100kpc.
\end{abstract}

\section{INTRODUCTION}

Blazars are a radio-loud subclass of active galactic nuclei, where the relativistic jet points directly at Earth. Blazars are characterized by two broad components in their spectral energy distribution (SED). The first one peaking between IR and X-ray energies is attributed to synchrotron emission of highly relativistic electrons, while the second component, which peaks between MeV and TeV $\gamma$-ray energies, is attributed to inverse Compton emission of the same electron population. We will concentrate here on this so-called leptonic emission model, which neglects possible contributions by hadronic interactions. Blazars can be divided into two main categories depending on whether or not they exhibit broad optical emission lines ($\Delta EW >5$\AA). BL Lac objects are generally devoid of broad emission lines. They may be further characterized by the peak position of their synchrotron component in the SED. Low-energy peaked BL Lac objects (LBLs) exhibit the synchrotron maximum below $10^{14}$Hz, while intermediate-energy peaked BL Lac objects peak between $10^{14}$Hz and $10^{15}$Hz, and high-energy peaked BL Lac objects peak above $10^{15}$Hz. 

The blazar AP Librae, located at a redshift $z=0.0486$ and at $\mbox{R.A.}_{\mbox{J2000}} = 15^h 17^m 41.8^s$, $\mbox{DEC.}_{\mbox{J2000}} = -24^{\circ} 22^{\prime} 19.5^{\prime\prime}$, exhibits a monotonically increasing X-ray energy spectrum. This in combination with the lack of optical emission lines has led to its classification as an LBL. Interestingly, some observational features of this source do not fit into this category.

Since the synchrotron component of LBLs peaks below $10^{14}$Hz and the X-rays are produced by the inverse Compton (IC) process, the maximum electron Lorentz factor in the electron energy distribution cannot significantly exceed $\sim 10^4$ for reasonable values of the magnetic field strength on the order of $\sim 0.1$G. Hence, one would not expect very high energy $\gamma$-ray (VHE, $E>100$GeV) emission from LBLs. Surprisingly, AP Librae has been clearly detected by observations with the H.E.S.S. telescope array \cite{hea15}, and the SED extends to energies of a few TeV. Currently, AP Librae is the only LBL listed in the TeVCat\footnote{\url{http://tevcat.uchicago.edu/}}, a catalog that gathers all sources detected above $100$GeV. Despite selection biases, this makes AP Librae an exceptional source.

High resolution X-ray observations led to the detection of extended X-ray jets in many AGN. However, due to the small viewing angle, it is surprising to observe extended X-ray emission in blazars. An extended X-ray jet has been detected in AP Librae by \cite{kwt13}, which has thus become one of only six BL Lac objects listed in the X-JET database.\footnote{\url{http://hea-www.cfa.harvard.edu/XJET/}} Of these six objects three exhibit a synchrotron dominated X-ray spectrum, while the other three, including AP Librae, are IC dominated. The X-ray morphology of AP Librae's jet follows exactly the radio morphology as observed with the VLA. The detection of the extended X-ray jet further demonstrates AP Librae's peculiarity.

The detection in VHE implies an extremely broad high energetic component spanning more than 10 orders of magnitude in energy. Since the synchrotron emission cuts off below the X-ray band, the resulting narrow electron distribution cannot explain the broad IC component in the usual one-zone blazar model. 

In this proceeding, we summarize our recent result \cite{zw15} that the extended jet dominates the total SED in the VHE $\gamma$-ray regime. Our model explains the VHE emission as originating mostly from the extended jet. 

\section{IMPORTANT OBSERVATIONS}

We describe only the important observations, which are necessary for our jet model. The remaining multiwavelength data is taken from the following papers or publicly available data bases: Radio \cite{kea81, pla1}, and IR/optical/UV \cite{kwt13, hbs15}.

\subsection{Radio}

\subsubsection{VLA}

VLA observations \cite{cea99} at $1.36\,$GHz led to the detection of the extended jet on arcsec-scales emerging in a south-westerly direction. Beyond $\sim 12\,$arcsec the jet bends to the north-west for another $\sim 10-20\,$arcsec. The spectral point of the extended radio jet is marked by the blue square at $1.36\,$GHz.

\subsubsection{MOJAVE}

The MOJAVE program \cite{lea09} utilizes VLBI radio observations at $15\,$GHz to monitor blazars on milli-arcsec scales over long time periods. AP Librae was observed in this program over the course of $\sim 15$ years. The data set reveals a steady core component (``component 0''), which is weakly variable (flux within a factor of 2). Its flux is marked with the open diamond in Fig. \ref{fig:sed}. 

Furthermore, a continuous jet is measured on scales of $\sim 10$milli-arcsec, which emerges in a southerly direction. Beyond this scale only knots in a south-westerly direction are detected. This is the same position angle as seen for the arcsec scale jet from the VLA observations. The total flux (containing both the component 0 and the extended flux) obtained from the MOJAVE data fits with the spectrum of the non-VLBI radio data (c.f. Fig. \ref{fig:sed}). Hence, these data points are influenced by the extended component proving a usually made assumption of radio flux points of blazars.

The movement of the fastest knots led to the determination of a maximum apparent speed of the jet of $6.8c$.

\subsection{X-rays}

X-ray observations with Chandra revealed the extended jet on arcsec scales \cite{kwt13}. The photon index of the jet is $\Gamma = 1.8\pm 0.1$; thus indicating an IC dominance in this spectral regime. The spectral points are marked by blue squares in Figure \ref{fig:sed}. The morphological structure is very similar to the radio morphology of the VLA observation.

The Chandra spectrum of the core is also hard with $\Gamma = 1.58\pm 0.04$. 

A 100-month average of observations with the Swift-BAT instrument \cite{pal100} reveals a flux level of the hard X-rays that is in straight extrapolation of the Chandra core spectrum. Thus, the core X-ray spectrum can be described by a single power-law over more than 2 orders of magnitude in energy.

\subsection{$\gamma$-rays}

Due to a lack of spatial resolution of the $\gamma$-ray instruments, the jet cannot be resolved. Data from Fermi and H.E.S.S. are taken from \cite{hea15}. The $\gamma$-ray SED can be described by a flat level below a few GeV, followed by a power-law with larger index up to a few TeV. There is no apparent cut-off in the H.E.S.S. spectrum.

\section{THE JET MODEL}

\begin{figure}
\centering{\includegraphics[width=0.8\textwidth]{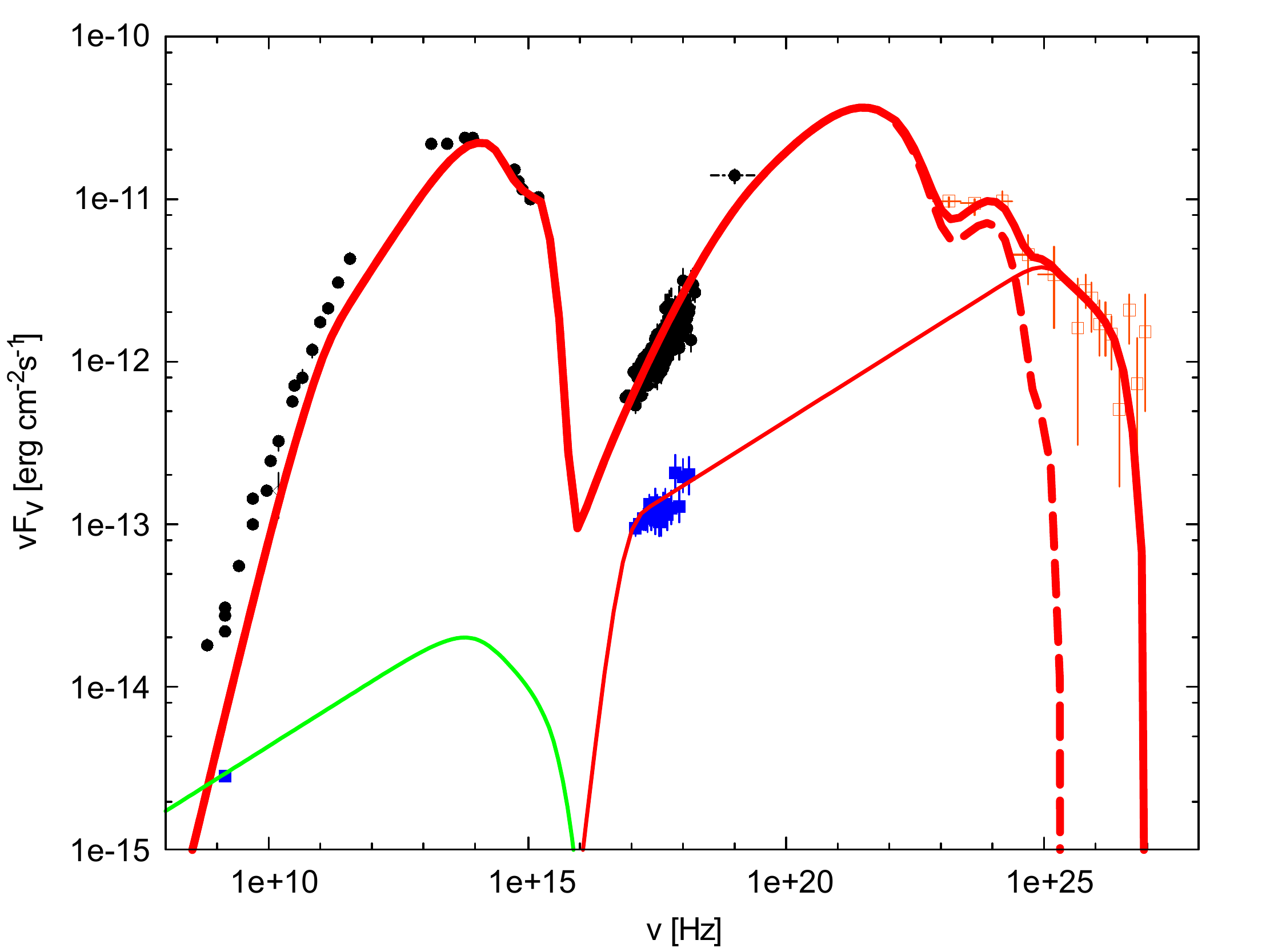}}
\caption{SED of AP Librae with the modeling of the blob and the kpc-scale jet. The data points are for the core (black dots), the kpc-scale jet (blue squares), the steady component of the MOJAVE observation (open diamond), and the $\gamma$-ray data (red squares) where the jet cannot be resolved. The line styles refer to the blob (thick dashed red), and the kpc-scale jet (thin solid lines). The line colors of the jet model imply synchrotron (green), and IC/CMB (red) emission. The thick red line is the sum of the blob and the jet.}
\label{fig:sed}
\end{figure} 
\begin{table}
\begin{tabular}{lcc}
				  & blob 		& kpc-scale jet \\
	\hline
	$n_e$ [cm$^{-3}$] 	  & $2.9\times 10^{2}$	& $1.6\times 10^{-9}$ \\
	$\gamma_{min}$ 		  & $1.6\times 10^2$ 	& $4.0\times 10^1$ \\
	$\gamma_{br}$ 		  & $9.1\times 10^3$ 	& $6.2\times 10^5$ \\
	$\gamma_{max}$ 		  & $1.0\times 10^4$ 	& $5.0\times 10^6$ \\
	$s_1$ 			  & $2.0$ 		& $2.6$ \\
	$B_{0}$ [G] 		  & $0.04$ 		& $2.3\times 10^{-6}$ \\
	$\Gamma_{b}$ 		  & $10$ 		& $10$ \\
	$R$ [cm] 		  & $9.0\times 10^{15}$ & $1.5\times 10^{21}$ \\
	$\vartheta_{obs}$ [deg]	  & $2.0$ 		& $4.0$ \\
	$L_{disk}$ [erg/s] 	  & $1.3\times 10^{44}$	& - \\
	$L_{torus}$ [erg/s] 	  & $4.0\times 10^{42}$	& - \\
	$l_{jet}^{\prime}$ [pc]	  & - 			& $1.0\times 10^{4}$ \\
\end{tabular}
\caption{Parameters for the fit. $n_e$ is the electron density; $\gamma_{min}$, $\gamma_{br}$ and $\gamma_{max}$ are the minimum, break and maximum electron Lorentz factor, respectively; $s_1$ is the electron spectral index below the break; $B_0$ is the magnetic field strength; $\Gamma_b$ is the bulk Lorentz factor; $R$ is the radius of the components; $\vartheta_{obs}$ is the observation angle; $L_{disk}$ and $L_{torus}$ are the luminosity of the disk and the dusty torus, respectively; $l_{jet}^{\prime}$ is the observed jet length.}
\label{tab:inputcom}
\end{table}

The data together with the modeling is presented in Figure \ref{fig:sed}. The respective parameter sets are given in Table \ref{tab:inputcom}.

Blazars are commonly described by a one-zone model, which in most cases gives successful fits to the SED. This model fails for AP Librae. The most important constraint comes from the cut-off in the synchrotron component at UV energies and the fact that the Swift-BAT flux point is slightly above the $\gamma$-ray flux level. The latter implies that the maximum of the IC component must be located between $100\,$keV and $100\,$MeV. Both constraints require a maximum electron Lorentz factor $\gamma_2\lesssim 10^4$. From below the electron distribution is constrained by the hard X-ray spectrum. Attributing the X-ray core data to synchrotron-self Compton (SSC) emission requires a high minimum electron Lorentz factor of $\gtrsim 100$, because the SSC spectrum is a pure power-law only below the photon energy associated with the minimum electron energy. Above this energy the SSC spectrum is significantly curved due to the convolution of the broken electron distribution with the synchrotron spectrum and the complicated IC cross section. Thus, the Chandra and the Swift-BAT spectrum cannot be fitted if the minimum Lorentz factor is set below $100$.\footnote{We note that a successful fit of the X-ray spectrum is also possible with a hard electron distribution with spectral index $s=1.5$. However, the issue of the narrow electron distribution is not circumvented.}

In turn, the electron distribution function, which explains the synchrotron core component and the X-ray core data, is very narrow, and cannot account for the VHE $\gamma$-ray emission. The resulting model is presented with a thick dashed line in Fig. \ref{fig:sed}. The fit is good for most of the synchrotron component and fits the ``component 0'' flux point of the MOJAVE observations. The rest of the radio emission includes contributions from the milli-arcsecond jet. The model describing this data is not shown here for clarity, and can be found in \cite{zw15}. The flattening of the Swift-UVOT spectrum at higher UV energies suggest the addition of another component, which is usually attributed to the accretion disk. The accretion disk can illuminate the gas in close proximity of the black hole, which is interpreted here as a low-luminous torus. Both components, whose parameters are given in Table \ref{tab:inputcom}, are also not shown for clarity. These two photon fields can serve as target photons for IC scattering by the blob electrons. The resulting flux contributes in the $\gamma$-ray range, and can explain the flat flux level below a few GeV. However, due to the low maximum electron Lorentz factor, these emission processes cannot explain the VHE emission, either.

In order to explain the VHE $\gamma$-ray spectrum we consider the extended jet. It is modeled as the combination of a number of self-similar zones. The combined emission of all zones gives the total flux of the jet. 

The model of the kpc-scale jet is constrained by the VLA and Chandra data. Since the flux in both radio and X-rays drops beyond the bend \cite{kwt13}, we only consider the part closer than $\sim 12$arcsec from the core corresponding to a projected length of $\sim 10\,$kpc. The synchrotron and IC/CMB emission of the kpc-scale jet gives a good fit in the radio and the Chandra energy range. Interestingly, the extrapolation of the Chandra jet spectrum intersects the Fermi data roughly at the break at a few GeV. This led us to the hypothesis that the VHE emission could originate in the kpc-scale jet. In fact, by choosing a maximum electron Lorentz factor of $\sim 5\times 10^6$ in the kpc-jet, we successfully reproduce the VHE spectrum.

\section{DISCUSSION \& CONCLUSIONS}

As was shown in the previous section, the extended jet plays an important role in the radiative output of AP Librae. It is responsible for a significant part of the SED. 

Most importantly, the jet could be responsible for the VHE emission detected with the H.E.S.S. telescopes. This is an unusual interpretation, since it implies highly relativistic flows and continuous reacceleration of the jet material on very large scales, given that modeling gives a deprojected length of AP Librae's jet of $140$kpc. Interestingly, in any observations of any scales the counter-jet has been detected. Under the assumption that both jets are intrinsically identical, the counter-jet must be strongly de-beamed even at large distances from the galactic center. Thus, we can conclude that even at large distances the jet must exhibit relativistic speeds without significant deceleration. 

The reacceleration of particles could be achieved by small-scale turbulence within the jet. Shear-layer acceleration might also be a viable alternative, which could be acting on large scales \cite{lbs13}. Any of these possibilities would not disturb the observed homogeneity of the jet because they are of the size scale on the order of a supernova remnant and cannot be resolved at the distance of AP Librae.

Since the electrons emitting in the X-ray band are less energetic than the radio emitting electrons, the power in relativistic particles is high and in some jets exceeds the Eddington luminosity even without accounting for cold protons. Our modeling suggests that in AP Librae the Eddingtion luminosity constraint holds for the large-scale jet even if the dominating contribution of cold protons is included.

Below we present a few suggestions how this model could be tested. 

Due to the required highly relativistic electrons ($\gamma\gtrsim 10^6$), the jet should emit synchrotron emission up to the UV band. Since the contribution of the galaxy in the UV band is much reduced compared to the optical and IR bands, the jet should be detectable despite its proximity to the bright core. If the jet is not detected in the UV with an upper limit below the predicted flux of our model curve, the IC/CMB is ruled out as the origin of the VHE emission, because the required electron energies cannot be matched. 

Within the kpc-scale jet the relative flux decrease along the jet is expected to differ between the synchrotron and IC domains due to different beaming dependencies \cite{d95}. The current observations do not allow to quantify this effect. More sensitive mapping at radio and X-ray frequencies could test this prediction.

Observations at HE $\gamma$-ray energies have been used to rule out the IC/CMB model in a number of extended X-ray jets \cite{mg14,mea15}, since the model overpredicts the upper limits derived from Fermi observations. Given that our model fits perfectly the HE and VHE data of AP Librae, this test is not applicable here.

Nevertheless, the HE and VHE emission can be used to indirectly test the model. We do not expect variability in the extended jet emission, because such a large object requires to be steady over time scales much longer than a few times $l_{jet}/c$. Thus, the flux above a few GeV should not drop below the observed level. An increase in the flux at these energies during a flare might still originate from an additional flaring component and would not rule out the above model. However, such a component should also change the spectrum in the UV and potentially in the X-ray band.

Given that AP Librae's host is a bright elliptical galaxy, one can expect absorption of TeV photons by the galactic optical starlight, if the TeV photons originate from the central region. This effect is on the order of $\sim 6\%$ \cite{zcw16}, and thus below the spectral sensitivities of the current generation of Cherenkov experiments. However, the futur Cherenkov Telescope Array might have the required sensitivity to detect the small absorption feature. This could be used to narrow down the location of TeV emission region. In our jet model, no absorption of TeV emission within the host galaxy is expected.

More observations in all energy bands are strongly encouraged.

If confirmed, the emission of TeV radiation by an extended, more than $100$kpc long jet would be a major result with strong implications for the transport and acceleration processes on large spatial scales.

\section{ACKNOWLEDGMENTS}
The authors wish to thank Markus B\"ottcher for the numerical code, which is described in detail in \cite{bea13}.
Support by the German Ministry for Education and Research (BMBF) through Verbundforschung Astroteilchenphysik grant 05A11VH2 is gratefully acknowledged. 
This research has made use of data from the MOJAVE database that is maintained by the MOJAVE team \cite{lea09}. 
This paper is based on observations obtained with Planck (http://www.esa.int/Planck), an ESA science mission with instruments and contributions directly funded by ESA Member States, NASA, and Canada.


\end{document}